\input{aipcheck}

\documentclass[
    ,final            
  ]
  {aipproc}

\layoutstyle{6x9}

\def\gta{\ifmmode {\mathbin{\lower 3pt\hbox   
    {$\,\rlap{\raise 5pt\hbox{$\char'076$}}\mathchar"7218\,$}}}
    \else {${\mathbin{\lower 3pt\hbox
    {$\rlap{\raise 5pt\hbox{$\char'076$}}\mathchar"7218\,$}}}
    $}\fi}
\def\lta{\ifmmode {\,\mathbin{\lower 3pt\hbox   
    {$\,\rlap{\raise 5pt\hbox{$\char'074$}}\mathchar"7218\,$}}}
    \else {${\mathbin{\lower 3pt\hbox
    {$\rlap{\raise 5pt\hbox{$\char'074$}}\mathchar"7218\,$}}}
    $}\fi}

\begin{document}

\title{Three-Body Encounters of Black Holes in Globular Clusters}

\author{Kayhan G\"{u}ltekin, \ M. Coleman Miller \ and Douglas P. Hamilton}{
  address={University of Maryland, Department of Astronomy,
College Park, MD 20742--2421}
}

\begin{abstract}
  Evidence has been mounting for the existence of black holes with
  masses from $10^{2}$ to $10^{4}\;M_{\odot}$ associated with stellar
  clusters.  Such intermediate-mass black holes (IMBHs) will encounter
  other black holes in the dense cores of these clusters.  The
  binaries produced in these interactions will be perturbed by other
  objects as well thus changing the orbital characteristics of the
  binaries.  These binaries and their subsequent mergers due to
  gravitational radiation are important sources of gravitational
  waves.  We present the results of numerical simulations of high mass
  ratio encounters, which help clarify the interactions of
  intermediate-mass black holes in globular clusters and help
  determine what types of detectable gravitational wave signatures are
  likely.
\end{abstract}

\maketitle


\section{Introduction}

Recent observations have given rise to the possibility of large black
holes located in the centers of stellar clusters.  Optical
observations of velocity profiles of M15 and G1 are consistent with
$2.5\times10^{3}$ and $2.0\times10^{4}\;M_{\odot}$ black holes (BHs)
(Gebhardt et al. 2000; Gerssen et al. 2002; van der Marel et al. 2002;
Gebhardt, Rich, \& Ho 2002; for a review see van der Marel, this
volume and for other interpretations see Baumgardt et al. 2003). X-ray
observations show unresolved, non-nuclear sources associated with both
young and globular clusters with $L \approx 10^{39}$ to
$10^{41}$~erg$\;$s$^{-1}$ in multiple galaxies; for a review see
Mushotzky, this volume.  The observed variability and fluxes, if
neither beamed nor super-Eddington, indicate BHs with M$\sim
10^{3}$~$M_{\odot}$.

The existence of IMBHs suggests a formation mechanism different from
those of stellar-mass BHs and supermassive BHs.  Several models have
been proposed to account for the origin of IMBHs including formation
from population III stars (Madau \& Rees 2001; Schneider et al. 2002),
interactions of stars in young clusters, and interactions of compact
objects in old clusters (Miller \& Hamilton 2002; Portegies Zwart \&
McMillan 2002; for a review see Miller, this volume).

Wherever and however IMBHs formed, the best candidates are found in
clusters where three-body encounters are important.  Any IMBH in the
center of a stellar cluster will pick up companions and undergo
three-body encounters.  These binaries and their mergers are important
sources of gravitational waves.  Advanced LIGO is expected to detect
the merger, and LISA is expected to detect the inspiral phase.  In
order to predict the gravitational wave signature of the merger, the
expected separations and eccentricities of the binaries must be known.
As three-body encounters alter these quantities, simulations of these
encounters are needed to predict their distributions.  In addition,
the simulations are useful for estimating the source population and
event rates.

\section{Numerical Simulations}

We perform numerical simulations of a series of three-body encounters
between high mass-ratio binary point masses and an interloping point
mass in Newtonian gravity.  We simulate an encounter between a hard
binary and an interloper and then use the resulting binary for the
next encounter.  This is repeated until the binary merges due to
gravitational radiation before its next encounter.  This study of a
sequence of encounters of high mass-ratio binaries differs from
previous studies of three-body encounters, which have focused on
nearly equal masses and single encounters.

Simulations were integrated using HNBody (Rauch \& Hamilton 2003).
The results presented here include pure Newtonian integrations of two
mass-ratios: 10000 sequences of 100:10:10
(dominant:companion:interloper) and 3000 sequences of 1000:10:10.
Both have initially circular orbits with a separation of $a = $10~AU.
A Monte Carlo initial condition generator samples all incoming
directions and orientations for significant encounters.  We also present 
simulations, described below, with general relativistic effects.

\section{Results of Newtonian Simulations}

\begin{figure}
  \rotatebox{90}{\includegraphics[height=.4\textheight]{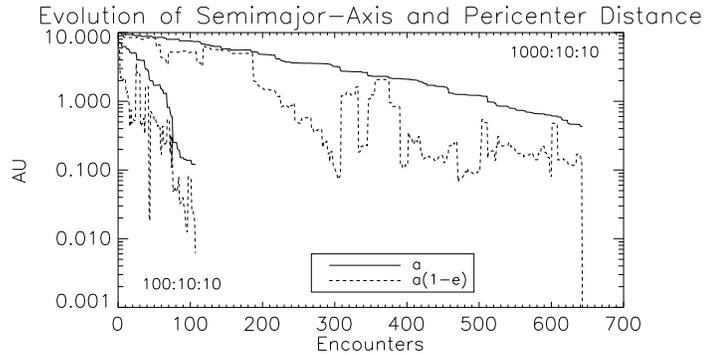}}
  \caption{Semimajor axis (solid) and pericenter distance (dashed) as a 
    function of number of encounters for 100:10:10 mass-ratio (left) 
    and 1000:10:10 (right).  As the binary undergoes a sequence 
    of three-body interactions, its semimajor axis decreases steadily, 
    but the eccentricity, and thus the pericenter distance, jumps from 
    low to high values in a single encounter.  The 100:10:10 mass-ratio 
    decreases in many fewer interactions because the interactions are 
    stronger.
    }
  \label{lifetime}
\end{figure}
Figure~\ref{lifetime} shows the changes in $a$ and $a(1-e)$ with each
encounter for two typical sequences.  For both mass-ratios, the
semimajor axis changes steadily, but the eccentricity and, thus, the
pericenter distance change drastically.  Both binaries merge at high
$e$ (see Fig.~2a) because the time to merge by gravitational radiation
is
\begin{equation}
  \tau \approx 3\times10^{8} 
  \left(M_{\odot}^{3}/\mu_{\rm bin} M_{\rm bin}^2\right)
  \left(a/R_{\odot}\right)^{4} \left(1-e^{2}\right)^{7/2}\;{\rm yr},
\label{grtime}
\end{equation}
where $M_{\rm bin}$ and $\mu_{\rm bin}$ are the mass and reduced mass
of the binary (Peter 1964).  The lower mass-ratio merges with fewer encounters
because the energy that the interloper can carry away scales as
$\Delta E \sim \left(m_{\rm comp}/M_{\rm bin}\right)$, where $m_{\rm
  comp}$ is the less massive binary member (Quinlan 1996).

\begin{figure}
  \begin{minipage}[t]{2.9truein}
    \mbox{}\\
    \rotatebox{90}{\includegraphics[height=.3\textheight]{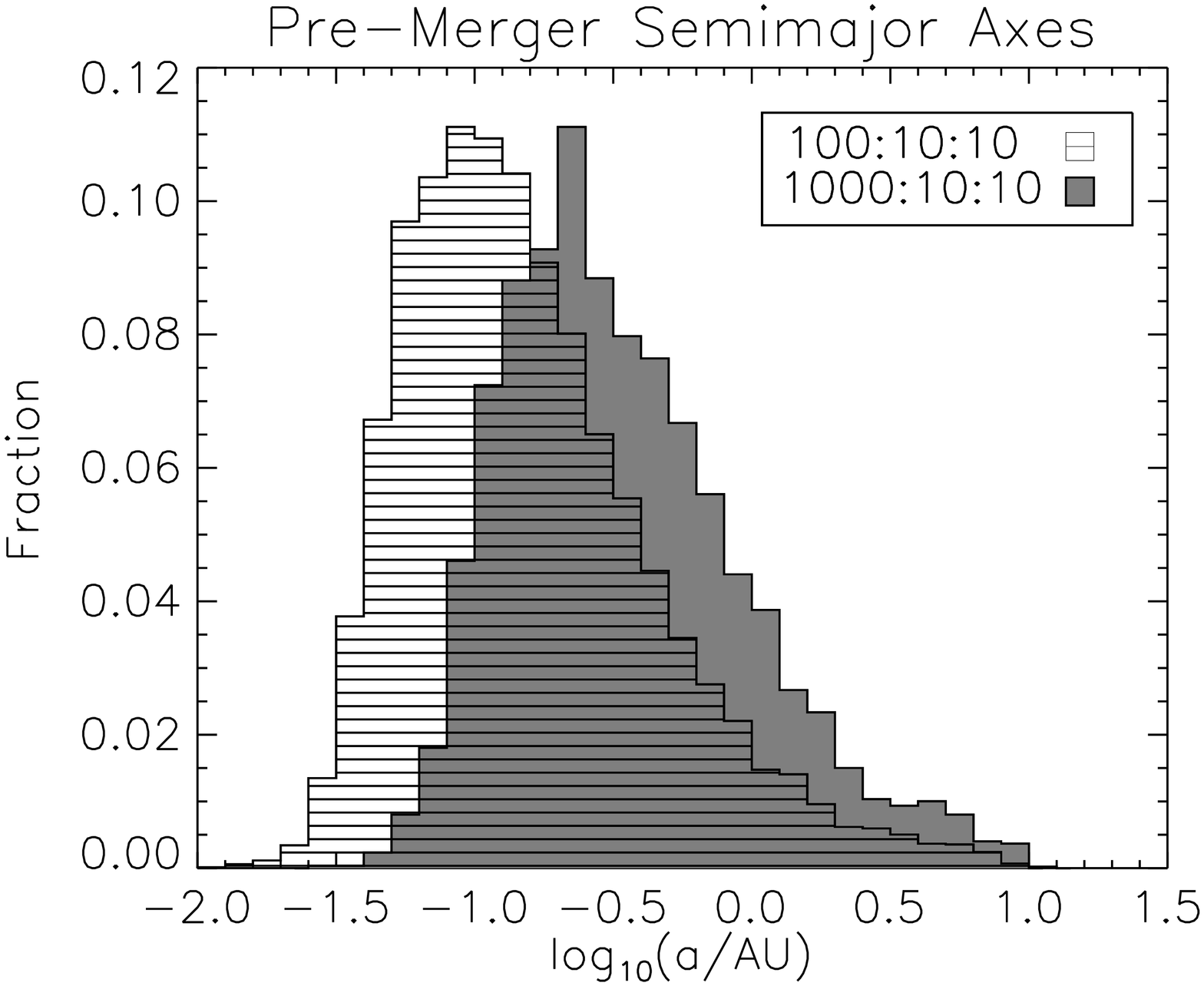}}
  \end{minipage}
  \hfill
  \begin{minipage}[t]{3.0truein}
    \mbox{}\\
    \rotatebox{90}{\includegraphics[height=.3\textheight]{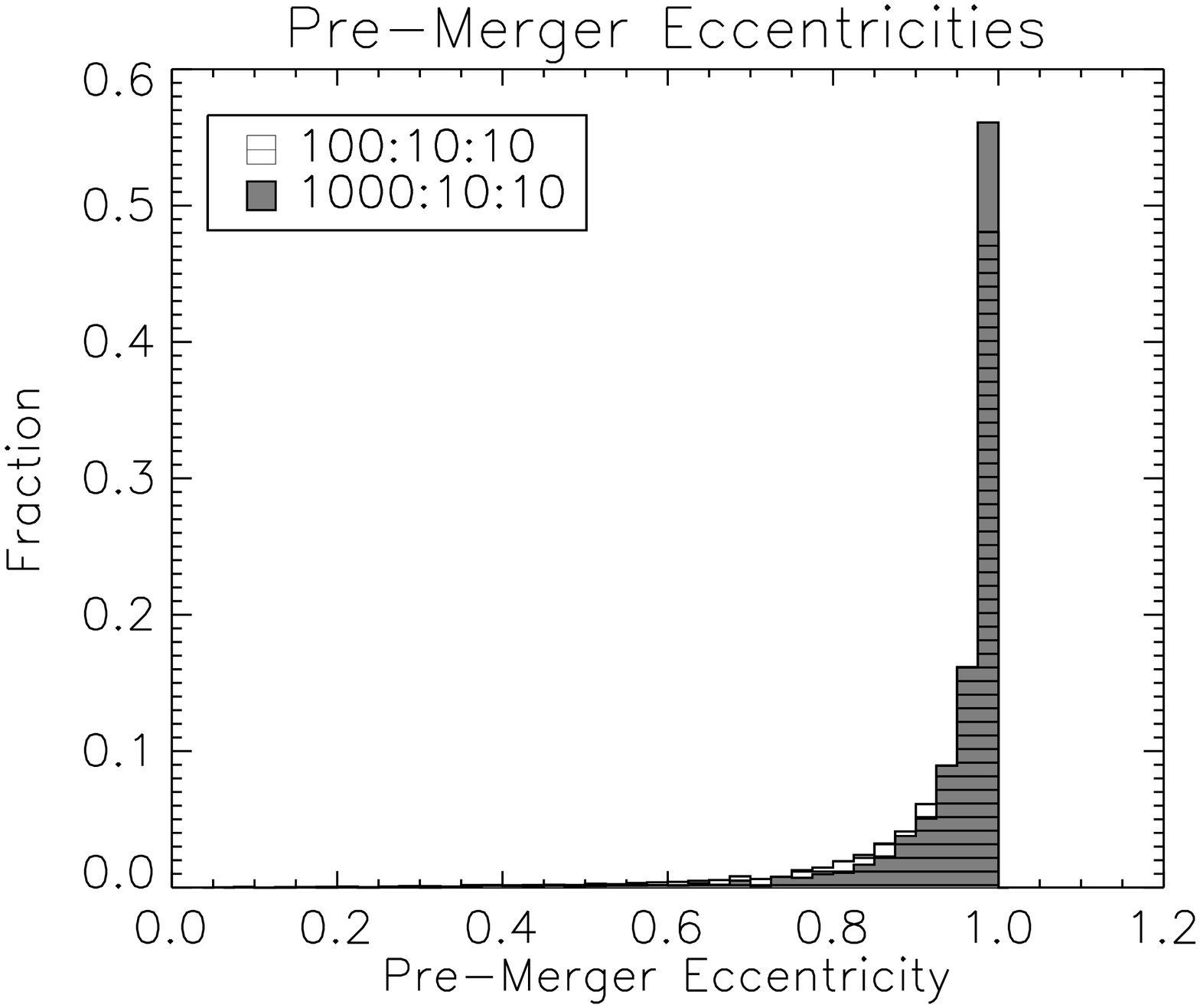}}
  \end{minipage}
  \caption{
    The left panel shows a histogram of the binary's semimajor axis
    after its last encounter and before it begins to merge.  The
    100:10:10 sequence (hatched) is shifted to a lower $a$ than the
    1000:10:10 sequence (solid) because the binaries with small $a$
    will tend to merge before undergoing another encounter.  The right
    panel shows a histogram of the binary's pre-merger eccentricity.
    The 100:10:10 sequence (hatched) and 1000:10:10 sequence (solid)
    are very similar in shape, and they are both peaked at $e\approx1$
    because of merger time's strong dependence on $e$ (see
    Eqn.~\ref{grtime}) and because $e$ can change from a low value to
    nearly unity in one strong interaction (see Fig.~\ref{lifetime}).
  }
\end{figure}

Figure~2a shows a histogram of the binary's semimajor axis after its
last encounter. The binary will then merge before it encounters
another object.  The mean semimajor axis is 0.32~AU for 100:10:10 and
0.64~AU for 1000:10:10.  Both mass-ratios have a similar shape whose
drop off at low $a$ is because the binary tends to merge before its
semimajor axis can decrease further.  The lower mass-ratio's histogram
is shifted to lower $a$ because the less massive binary will emit less
gravitational radiation, and hence shrink less, for a given orbit.

Figure~2b shows a histogram of the binary's eccentricity after the
last encounter.  The mean value is 0.930 for 100:10:10 and 0.950 for
1000:10:10.  The histograms of both mass ratios have very similar
shapes and are strongly peaked near $e = 1$.  The high eccentricity
right before merger is due to the strong dependence of merger time
on eccentricity and because the eccentricity can change drastically in
one encounter while the semimajor axis tends to decrease at a roughly
constant rate.  A high eccentricity is important from a gravitational
wave detection standpoint because the waveform of the gravitational
radiation emitted by high eccentricity binaries at inspiral is
significantly different from that of circular binaries in LISA's band.

\section{Adding General Relativistic Effects}

To augment our Newtonian treatment of this problem, we modified the
integrations to incorporate the effects of gravitational radiation.  In
between encounters the binary emits gravitational radiation causing the
orbit to shrink and circularize.  We add this effect to our simulations for
several mass ratios to test how it changes our results.  We compare
300 pure Newtonian and 300 runs with this general relativistic effect
for each of three mass-ratios considered: 10:10:10, 100:10:10, and
1000:10:10$\;M_{\odot}$.  We are also beginning simulations that 
include gravitational radiation during the encounter.

\begin{table}
\begin{tabular}{rrrrrrrrr}
\hline
  \tablehead{1}{r}{b}{Mass\\Ratio} &
  \tablehead{2}{c}{b}{Number of\\Encounters} &
  \tablehead{2}{c}{b}{Pre-merger\\$a$/AU} &
  \tablehead{2}{c}{b}{Pre-merger\\$e$} &
  \tablehead{2}{c}{b}{Interloper\\Ejections} \\

  & \tablehead{1}{r}{b}{Newt.}
  & \tablehead{1}{r}{b}{GR}
  & \tablehead{1}{r}{b}{Newt.}
  & \tablehead{1}{r}{b}{GR}
  & \tablehead{1}{r}{b}{Newt.} 
  & \tablehead{1}{r}{b}{GR}
  & \tablehead{1}{r}{b}{Newt.} 
  & \tablehead{1}{r}{b}{GR} \\
\hline
 10:10:10   &  52 &  48 & 0.17 & 0.17 & 0.92 & 0.89 &   8.7 &  7.2\\
 100:10:10  & 103 &  94 & 0.34 & 0.37 & 0.94 & 0.87 &  25.2 & 20.8\\
 1000:10:10 & 554 & 484 & 0.63 & 0.53 & 0.95 & 0.84 & 114.5 & 88.4\\
\hline
\end{tabular}
\caption{Simulations of binary evolution with two models:  (1) pure Newtonian gravity and (2) Newtonian gravity with gravitational radiation between encounters.  Typical variation in the average numbers in a run of 100 sequences is between 1\% (for pre-merger $e$) and 10\% (for pre-merger $a$).}
\label{grbinevol}
\end{table}

The main results are summarized in Table~\ref{grbinevol}.
Gravitational radiation shrinks and circularizes orbits, hence the
pre-merger $a$ and $e$, as well as the number of encounters and number
of interlopers ejected during the hardening, are all decreased because
of the radiation.  Thus general relativistic effects are important in
determining the orbital characteristics of the typical binary
inspiral.

We monitored the simulations for encounters that resulted in an
ejection of one of the less massive BHs from the globular cluster,
assuming an escape velocity of $50\;{\rm km\;s^{-1}}$.  This is
important in determining whether IMBHs can be built up from mergers of
stellar-mass BHs as proposed by Miller \& Hamilton (2002).  If the
build-up to the inferred masses requires more BHs than are available
in a cluster, the model cannot explain the formation of IMBHs.  Miller
\& Hamilton (2002) estimate $100\left(M_{\rm bin}/50 M_{\odot}\right)
\approx 2000$ ejections for a mass ratio of 1000:10:10, assuming an
eccentricity at merger of $e=0.7$.  This is far in excess of the 
$\sim 100$ ejections we find because the binaries will merge with an 
eccentricity much higher than the average of a thermal distribution.  
Thus merging stellar-mass BHs in a globular cluster may be more 
efficient than previously expected.

{\bf Acknowledgements}

This work was supported in part by NASA grant NAG 5-13229.

\end{document}